# Continuous frequency tuning of an external cavity diode laser significantly beyond the free spectral range by sweeping the injection current

**AKIFUMI TAKAMIZAWA**

*National Metrology Institute of Japan, National Institute of Advanced Industrial Science and Technology (AIST), 1-1-1 Umezono, Tsukuba, Ibaraki 305-8563, Japan*
*akifumi.takamizawa@aist.go.jp*

**Abstract:** In this study, the frequency of an external cavity diode laser with an antireflection-coated laser diode was continuously tuned over a range of 14.8 GHz, which was 4.5 times larger than the free spectral range, by only sweeping the injection current to the laser diode. Without the antireflection coating, the tuning range was reduced to one-fifth of the free spectral range, and mode hops occurred with hysteresis. The theoretical analysis starting from the hysteresis indicated that an unexpectedly wide tuning is possible if the longitudinal modes of the solitary laser diode vanish owing to antireflection coating.

## 1. Introduction

Tunable lasers with frequency control in the vicinity of atomic transitions are essential for conducting experiments such as laser cooling and atom manipulation in the field of atomic physics [1]. Recently, laser cooling of atoms has been applied to precision measurements, including clocks [2,3], magnetometers [4], and gravimeters [5,6], and the operation must be as stable as possible in these applications. One of the major causes for the unintentional stoppage of the operation is the mode hops of lasers, due to which a frequency lock to an atomic transition does not work. Therefore, decreasing the mode hops of lasers in the long term is important.

External cavity diode lasers (ECDLs) are the commonly used tunable lasers by virtue of their compactness and low cost [7,8]. Several types of ECDLs with optical components for wavelength selection, such as reflective grating [8–14], Bragg grating [15–17], microresonators [18], and interference filters (IFs) [10,19–24], have been developed. Several studies have been conducted to improve the performance of ECDLs in terms of their short-term stability [9,10,15–18], long-term stability [11–13,16,17,19–23], and tunability [10,14,24]. Among these, long-term stability is essential for eliminating unintentional mode hops during operation. ECDLs with an IF inserted into the external cavity (IF-ECDL) for improving their long-term stability are structurally insensitive to misalignment because a cat's eye configuration, in which the laser light is focused on both ends of the external cavity, can be easily adopted [19]. However, all ECDLs except for reflective-grating ECDLs generally have the limitation of a relatively narrow continuous tuning range (CTR) because it is difficult to synchronize the wavelength selected by the optical component with the frequency of the longitudinal modes determined by the length of the external cavity. By contrast, in reflective-grating ECDLs, wherein grating plays the role of both an output coupler and a wavelength selection component, synchronization can be achieved by adapting a well-designed geometrical arrangement to move the grating. Consequently, continuous tuning over a range greater than 10 THz has been realized [14]. In conventional IF-ECDLs, the CTR is not strongly considered; however, their long-term stability is pursued as an advantage. Therefore, the IF does not move; that is, the selected wavelength remains unchanged when the laser frequency is scanned by actuating an output coupler. As a result, the CTR is limited to the free spectral range (FSR) of the external cavity, typically a few gigahertz, if the longitudinal mode with the largest gain is always selected in modes equally arranged with an interval of the FSR.

IF-ECDLs have also been developed as laser sources for atomic fountain primary frequency standards by our group [25], wherein continuous operation for 1 month or longer is required. To render them mechanically more robust, all the optical components are fixed and any position adjusters are removed [20]. The frequency was tuned by changing the injection current into a laser diode (LD), whereas, in the commonly used IF-ECDLs, the frequency is generally scanned by displacing an output coupler with a piezo actuator. In the ECDLs proposed in this study, although the CTR was degraded to one-fifth of the FSR of the external cavity when the output facet of the LD was not coated with an antireflection (AR) material, the unintentional mechanical variation in the length of the external cavity was strongly suppressed. As a result, the frequency variation under free running is dominated by atmospheric pressure [20] and is largely suppressed by installing the ECDL in a housing sealed from the outside [21]. Considering the frequency lock at an atomic transition, a high passive stability contributes to a robust frequency lock, whereas a small CTR corresponds to a narrow locking range. Therefore, the frequency lock is affected when the CTR is too narrow, although broadening the CTR is not a direct purpose. In actual use, the benefit provided by the mechanical robustness surpasses the defect owing to the small CTR.

Frequency tuning by sweeping the LD current results from variations in the refractive index and, thus, the optical length of the active layer of the LD. By using an uncoated LD, the longitudinal modes of a solitary LD influence the ECDL [26]. Furthermore, by changing the optical length of the active layer, the frequency of the longitudinal modes of a solitary LD varies significantly more than that of an external cavity because of the large discrepancy between the optical lengths of a solitary LD and an external cavity. Consequently, the CTR of an ECDL with an uncoated LD becomes narrow when the LD current is swept. Therefore, it is expected that the CTR would be improved to be equal to the FSR using an LD with an AR coating on the output facet to eliminate unnecessary solitary LD modes.

As a result, the CTR was increased by a factor of 4.5 times wider than the FSR in the present study. In addition, in other studies, a CTR of more than 10 times the FSR in an IF-ECDL [22] and a CTR slightly exceeding the FSR in a Bragg-grating ECDL [17] by sweeping the LD current have been reported, although no in-depth investigations regarding wide CTRs have been performed.

Hysteresis on mode hops is observed in an IF-ECDL with an uncoated LD when the frequency is scanned by sweeping the LD current. Hysteresis indicates that the initial dominant mode is selected even when the gain in the mode is lower than that in adjacent modes. If the longitudinal mode with the largest gain is not necessarily selected, the CTR can be wider than the FSR when unnecessary solitary LD modes are sufficiently eliminated using an AR-coated LD. For an LD without an external cavity, the hysteresis on the mode hops has been investigated [27] and explained through a nonlinear effect that suppressed the gain of the adjacent modes [28]. However, to the best of the author's knowledge, no studies have focused on hysteresis in an ECDL.

In this study, the continuous frequency tuning of an IF-ECDL with an AR-coated LD significantly exceeding the FSR by only sweeping the LD current was performed. Moreover, the hysteresis of mode hops in an IF-ECDL with an uncoated LD was experimentally investigated. To explain the unexpectedly wide CTR, from the observed hysteresis, the extent of the broadening of the CTR of the IF-ECDL with an AR-coated LD was estimated.

## 2. Experiment

Figure 1 shows the IF-ECDL used in the experiment, which is essentially identical to that described in [20]. The ECDL was composed of a Fabry–Perot-type LD, three lenses, a partial mirror (PM), and an IF. The output from the LD was collimated by the first lens, focused on the PM by the second lens, and collimated again by the third lens. The focal lengths were 4.51 mm, 18.40 mm, and 11.00 mm, respectively. An external cavity with a reflectance of 30 % was created between the rear facet of the LD and the reflection surface of the PM. The setup was intentionally designed to achieve the given reflectance of the PM, whereas the reflectance of the rear facet of the LD was fortuitously found to be approximately the same (see Section 3). The IF used was a bandpass filter (BPF) with a transparence bandwidth of 0.3 nm at full width at half maximum (FWHM). The BPF was inserted between the first and second lenses. The central wavelength of the BPF depends on the tilt of the laser beam; therefore, the BPF was designed such that the central wavelength was tuned to 895 nm, which is the resonant wavelength of $^{133}$Cs D1 line, at an incident angle of 6°. The tilt of the BPF was adjusted such that the wavelength was near the atomic transition point. The optical length of the external cavity in terms of the refractive index of the vacuum was 46 mm, from which the FSR was calculated to be 3.3 GHz.

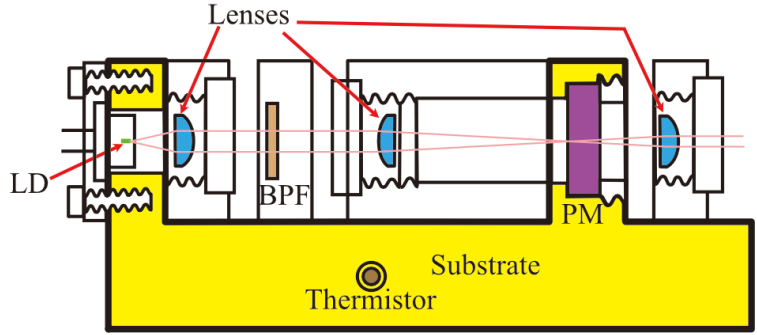

Fig. 1. Schematic illustration of the side view of the IF-ECDL. The yellow part represents the substrate that is sterically machined.

The LD with a can package and the PM were directly fixed with screws on a substrate made of sterically machined copper. When the positions of the LD and PM are fixed, the length of the external cavity is determined. Jigs supporting the three lenses and BPF were glued onto the substrate after adjusting their locations. Once the ECDL is fabricated, all the components are fixed and cannot be adjusted to tune the laser frequency. Therefore, the laser frequency was coarsely adjusted by controlling the temperature of the substrate using a Peltier element in which a thermistor was embedded into the substrate to monitor the temperature. Subsequently, the laser frequency was continuously tuned by varying the LD current.

Two IF-ECDLs were used in the experiment: (1) "ECDL-AR" with an AR-coated LD (Sacher, SAL-0920-060) and (2) "ECDL-UN" with an LD without AR coating (Thorlabs: M5-905-0100). Except for the coating on the LD facet, they are almost identical. According to the datasheet, the typical and specification reflectance values of the output facet of the AR-coated LD are $<5\times10^{-5}$ and $<5\times10^{-4}$, respectively. The supplier emphasized the process of evaporation of the coating; thus, the spectral fringes generated by the longitudinal modes were observed, and the thickness of the coating was adjusted to minimize the fringes [29]. The evaporation process might be important for the expansion of the CTR. Hence, by using an AR-coated LD with a similar specification of reflectance purchased from another supplier, the CTR of EDCL-AR was not significantly enlarged compared with that of ECDL-UN.

Additionally, there were two differences in the IF-ECDLs used in this study, which must have had a minor influence on continuous tuning. One difference was the housing: although the ECDL-UN was contained in a sealed housing to avoid the influence of variations in atmospheric pressure [21], the ECDL-AR was not. Sealed housing was considered unimportant in this study, as long-term measurements were not conducted. The other difference was the current controller for LDs: owing to the difference in noise, the spectral linewidths were 200 kHz for ECDL-AR and 40 kHz for ECDL-UN.

To measure the variation in the frequency of ECDL-AR (ECDL-UN), a beat signal was obtained using ECDL-UN (ECDL-AR) as a reference laser. The frequency of the beat signal was measured using a frequency counter. Figure 2 shows the frequency variation in ECDL-AR as a function of the LD current. The laser frequency was continuously tuned over a range of 14.8 GHz, which was 4.5 times larger than the FSR, at a rate of −0.12 GHz/mA. Notably, the minimum LD current providing the frequency data was close to the threshold current for laser emission (65 mA), whereas the maximum LD current was determined to avoid damage to the LD. No mode hops were observed over the entire range of the LD currents used in the experiment. No hysteresis was observed with respect to the LD current.

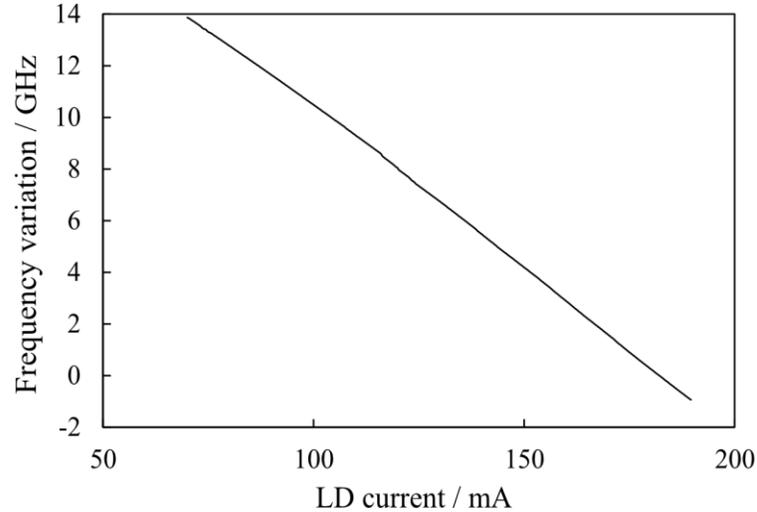

Fig. 2. Frequency variations in the ECDL-AR as a function of the LD current.

The spectrum of the beat signal was observed using a spectrum analyzer. Figures 3(a) and (b) show the spectra at LD currents of 130.0 mA and 88.5 mA, respectively. In most LD current ranges, a single longitudinal mode was obtained, as shown in Fig. 3(a). However, particularly at approximately 88.5 mA and 167.5 mA, a few side modes were observed, as shown in Fig. 3(b). The power of the maximal side mode was 10.7 dB lower than that of the carrier mode. The modes were equally separated by 2.9 GHz, differing somewhat from the FSR. Even when the side modes were excited, the continuous tuning of the carrier frequency was maintained. The difference between the laser frequencies of the carriers at LD currents of 88.5 mA and 167.5 mA was 9.9 GHz. The power of the maximal side mode was reduced to 20 dB below the power of the carrier at LD currents of 88.5 – 2.0 mA and 88.5 + 3.1 mA as well as at LD currents of 167.5 – 1.3 mA and 167.5 + 2.3 mA. Notably, no problems were observed while obtaining a single mode near a specific atomic resonance; even if the side modes accidentally appeared when the carrier frequency was tuned to the desired frequency, a single-mode oscillation was obtained at that frequency by adjusting both the LD current and temperature. However, single-mode oscillations were not acquired over a wide range of 14.8 GHz, even at any temperature as far as the author tried.

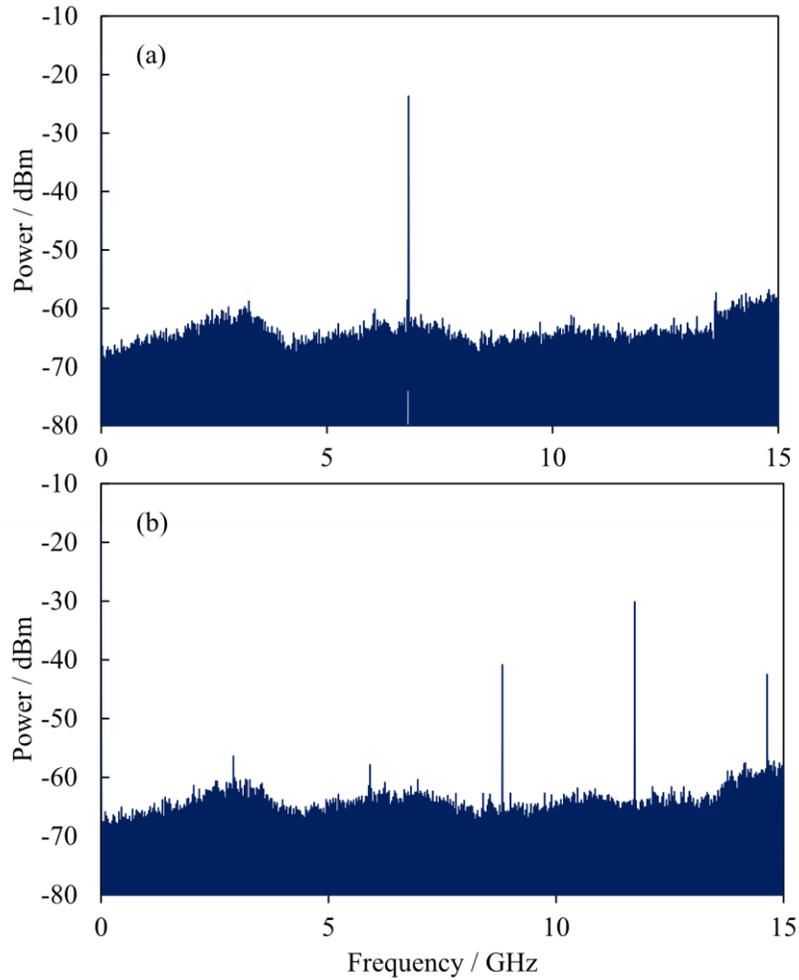

Fig. 3. Frequency spectra of the beat signals at LD currents of (a) 130.0 mA and (b) 88.5 mA in ECDL-AR.

Subsequently, the frequency of ECDL-UN was measured by increasing or decreasing the LD current. Figure 4 shows the variations in the laser frequency as a function of the LD current, where the blue and red lines indicate the cases in which the LD current increased and decreased, respectively. Compared with that of ECDL-AR, the CTR of ECDL-UN was considerably reduced and mode hops appeared. Subsequently, the hysteresis of the laser frequency was observed with respect to an increase and a decrease in the LD current. Notably, no mode hops occurred when the LD current was swept between C and F, as indicated in Fig. 4, once the ECDL mode was selected. Therefore, the CTR was 0.67 GHz in the frequency range between C and F. If no hysteresis had been observed, the CTR would have been 0.47 GHz, which is given by the frequency range between C/D and E/F, where C/D (E/F) is the midpoint of C and D (E and F). Therefore, the CTR increased by a factor of 1.4 owing to hysteresis.

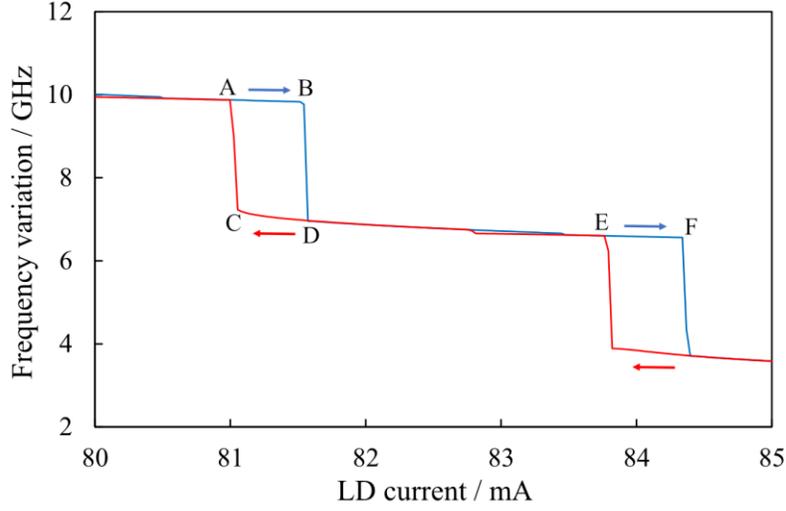

Fig. 4. Frequency variations of ECDL-UN as a function of the LD current, where the blue and red lines indicate the cases of increasing and decreasing the LD current, respectively. The symbols A–F represent the LD currents and frequencies at which mode hops occurred.

## 3. Discussion

To discuss the wide CTR of ECDL-AR, the hysteresis in ECDL-UN was first considered. Figure 5(a) schematically shows the transmittance spectra of the BPF, solitary LD mode, and ECDL modes, assuming the case of ECDL-UN. Here, the gain of the laser medium was neglected because the spectrum was considerably wider than that of the transmittance of the BPF and was thus regarded as homogeneous. In ECDL-UN, because the FSR of the solitary LD was approximately half the FWHM of the transparence bandwidth of the BPF, one of the solitary LD modes was selected by the BPF. When the LD current is decreased (increased), the frequencies of both solitary LD modes and ECDL modes shifted in the positive (negative) direction owing to variations in the refractive index of the active layer of the LD. Because of the small occupation of the active layer in the optical length of the external cavity, the frequency variation in solitary LD modes was 16 times faster than that of ECDL modes in ECDL-UN. Therefore, one mode hop occurs every time the frequencies of solitary LD modes shift by the FSR of ECDL modes.

The transmittance spectrum of solitary LD modes with a Fabry–Perot cavity is given as,

$$S_{\mathrm{LD}}(\nu) = \frac{(1-R_s)^2}{1+R_s^2 - 2R_s \cos(4\pi \nu n_s L_s/c)} \quad (1)$$

where $R_s$ is the reflectance of the facets at both ends; and $\nu$, $n_s$, $L_s$, and $c$ are the frequency, the refractive index of an active layer relative to vacuum, the cavity length of a solitary LD, and the speed of light in vacuum, respectively. Assuming that $n_s = 3.5$ and the LD facets are uncoated, $R_s = 0.31$. Moreover, $n_s L_s = 2.9$ mm by measuring the FSR of the same type of LD used in ECDL-UN. The FWs of the solitary LD modes were calculated using Eq. (1) to be 16 GHz, which is one-seventh of the FWHM of the BPF. For simplicity, the transmittance spectrum of the BPF was regarded as homogeneous when considering the ECDL-UN.

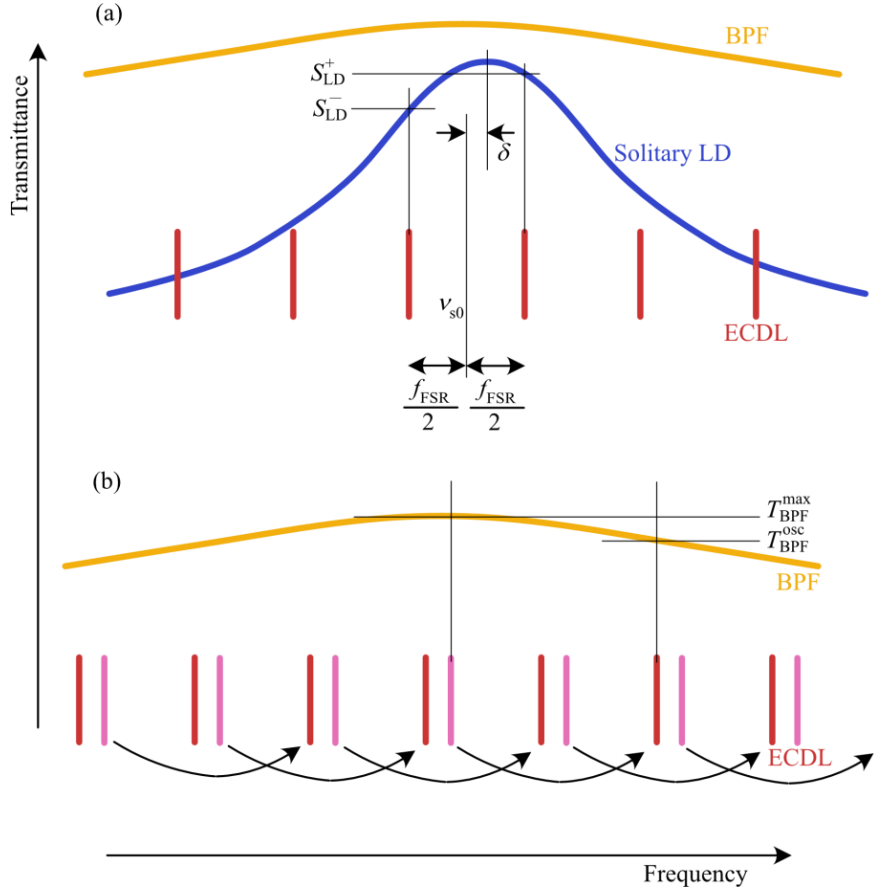

Fig. 5. Schematic illustration of the transmittance of the BPF, solitary LD mode, and ECDL modes in the cases of (a) ECDL-UN after the LD current is decreased, and (b) ECDL-AR. In (b) the pink lines represent the initial ECDL modes, and red lines represent the ECDL modes after the LD current is decreased.

Initially, it was assumed that the midpoint of the two successive ECDL modes matched the center of the solitary LD mode, whose frequency was denoted as $v_{s0}$. In other words, the frequencies of the two ECDL modes are $v_{s0} \pm f_{\text{FSR}}/2$, where $f_{\text{FSR}}$ is the FSR of the external cavity. Here, the transmittances of the solitary LD mode at ECDL modes on the positive and negative sides are denoted as $S_{\text{LD}}^+$ and $S_{\text{LD}}^-$, respectively, and the larger values of $S_{\text{LD}}^+$ and $S_{\text{LD}}^-$ are denoted as $S_{\text{LD}}^L$. When the LD current is decreased (increased), $S_{\text{LD}}^+$ ($S_{\text{LD}}^-$) becomes larger than $S_{\text{LD}}^-$ ($S_{\text{LD}}^+$) because the solitary LD modes shift in the positive (negative) direction significantly faster than ECDL modes, as shown in Fig. 5(a). When the central frequency of the solitary LD mode changes to $v_{s0} + \delta$ by varying the LD current, the proportion of the difference between $S_{\text{LD}}^+$ and $S_{\text{LD}}^-$ to $S_{\text{LD}}^L$, $\alpha \equiv (S_{\text{LD}}^+ - S_{\text{LD}}^-)/S_{\text{LD}}^L$ is given as,

$$\alpha(\delta) \simeq \left[ S_{\text{LD}}\left(\frac{f_{\text{FSR}}}{2} - \delta\right) - S_{\text{LD}}\left(-\frac{f_{\text{FSR}}}{2} - \delta\right) \right] / S_{\text{LD}}^L(\delta), \quad (2)$$

where

$$S_{\text{LD}}^L(\delta) \simeq \begin{cases} S_{\text{LD}}\left(\frac{f_{\text{FSR}}}{2} - \delta\right) & \text{at } \delta \geq 0 \\ S_{\text{LD}}\left(-\frac{f_{\text{FSR}}}{2} - \delta\right) & \text{at } \delta < 0 \end{cases}, \quad (3)$$

using the periodicity of $S_{LD}(v)$. Here, the frequency shift of ECDL modes was neglected compared with that of solitary LD mode. If hysteresis does not appear, the ECDL mode on the positive (negative) side is selected as an oscillation mode when $α(δ) > 0 (< 0)$. By contrast, if hysteresis occurs, the relationship is broken.

Because the refractive index of the active layer of a solitary LD varies linearly with the LD current [30], the optical lengths of the active layer and external cavity change linearly with the LD current. Although the frequency variation of ECDL-UN under continuous tuning was in nonlinearity with the LD current, as shown in Fig. 4, the discrepancy may be attributed to the frequency pulling induced by the solitary LD mode and BPF. The hysteresis frequency range, defined as the range of the central frequency of the solitary LD mode where the relation between the oscillation mode and the sign of $α(δ)$ is broken as described above, is given from the experimental result as follows:

$$\Delta f_h = \frac{\Delta I_h}{\Delta I_{FSR}} f_{FSR}, \tag{4}$$

where $\Delta I_h$ and $\Delta I_{FSR}$ are the ranges of the LD current between A and B and between A and E in Fig. 4, respectively. Here, $\Delta f_h = 0.65$ GHz by substituting $\Delta I_h = 0.54$ mA, $\Delta I_{FSR} = 2.76$ mA, and $f_{FSR} = 3.3$ GHz. Hysteresis occurred when $|δ| < \Delta f_h/2$. Therefore, the condition of $α(δ)$ in which the hysteresis is induced is derived as follows:

$$|α(δ)| < α\left(\frac{\Delta f_h}{2}\right) \equiv α_{max}. \tag{5}$$

In other words, once the ECDL mode on the positive (negative) side is selected as an oscillation mode under the condition of $α(δ) > 0 (< 0)$, it continues to be chosen even when $-α_{max} < α(δ) < 0$ ($0 < α(δ) < α_{max}$). By incorporating the value $δ = \Delta f_h/2$ into Eqs. (2) and (3), and using Eq. (1), $α_{max} \simeq 0.020$.

Next, the ECDL-AR was considered, wherein the solitary LD modes are negligible, as shown in Fig. 5(b). In this case, the ECDL mode was selected using the BPF. It was assumed that the frequency of one of the ECDL modes initially matches the center frequency of the transmittance of the BPF and that laser oscillation occurs in the central ECDL mode. When the LD current is decreased, all ECDL modes shift in the positive direction, whereas the transmittance spectrum of the BPF is fixed. Therefore, although the transmittance of the BPF in the oscillation mode decreases, that of the ECDL mode on the negative side increases. When the LD current is increased, the opposite occurs. Applying the discussion regarding ECDL-UN to ECDL-AR, the oscillation mode was permitted to sustain in the case where $β \equiv (T_{BPF}^{max} - T_{BPF}^{osc})/T_{BPF}^{max} < α_{max}$, where $T_{BPF}^{max}$ and $T_{BPF}^{osc}$ are the transmittance of the BPF at its maximum and that at the oscillation mode, respectively. From the transmittance spectrum of the BPF used in the experiment, the frequency range satisfying $β < α_{max}$ was estimated to be 16 GHz. Therefore, the experimental results of continuous frequency tuning over a range of 14.8 GHz significantly exceeding the FSR of the external cavity can be explained.

In the discussion above, $α_{max}$ was derived not from a theoretical analysis but from the experimentally obtained hysteresis frequency range. Here, $α_{max}$ may depend on the conditions of the specific ECDL. Therefore, it is difficult to adapt the discussion to quantitatively estimate the enhancement of the CTR of a general ECDL. For the generalization, a deeper theoretical analysis on hysteresis in an ECDL, which has not yet been focused on, will be needed.

## 4. Conclusion

In contrast to the CTR of conventional tunable lasers, in this study, by only sweeping the LD current, the CTR of the IF-ECDL was determined to be 4.5 times wider than the FSR of the external cavity by using an AR-coated LD to eliminate the solitary LD modes. However, when using an uncoated LD, hysteresis was observed in mode hops. Theoretical analysis starting from hysteresis explains the unexpectedly wide CTR of the IF-ECDL with an AR-coated LD.

In previous experiments [20,21], although the ECDLs were mechanically robust because of the lack of position adjusters, the CTR was considerably narrower than the FSR when using an uncoated LD. However, the present study resolves this limitation. The wide CTR of 14.8 GHz obtained herein is sufficient for a frequency lock under normal environmental conditions. For example, the CTR covers an atmospheric pressure range of 222 hPa according to the ratio between the laser frequency and the atmospheric pressure [20]. ECDLs with both mechanical robustness and wide CTRs are expected to be used for applications based on atomic physics, wherein lasers must be steadily frequency-locked to atomic transitions over the long term.

**Acknowledgements.** This study was supported by Innovative Science and Technology Initiative for Security Grant No. JPJ004596, ATLA, Japan. The author sincerely thanks S. Yanagimachi for his help in creating the setup for the data acquisition of the beat signal.